\documentclass[prd,aps,
nofootinbib,%
twocolumn
]{revtex4-1}

\usepackage{graphicx}
\usepackage{amsmath}
\usepackage{amssymb}
\usepackage{hyperref} 
\usepackage{color} 
\newcommand{\1}{}
\newcommand{\0}{}

\newcommand{\be}{\begin{equation}}
\newcommand{\ee}{\end{equation}}
\newcommand{\bea}{\begin{eqnarray}}
\newcommand{\eea}{\end{eqnarray}}
\newcommand{\nn}{\nonumber}
\newcommand{\6}{\partial }

\newcommand{\RSS}{R_{\rm D4}}
\newcommand{\MKK}{M_{\rm KK}}

\newcommand{\uKK}{u_{\rm KK}}
\newcommand{\gYM}{g_{\rm YM}}
\newcommand{\Nc}{N_{c}}
\newcommand{\ls}{l_{s}}
\newcommand{\gs}{g_{s}}

\newcommand{\Vv}{\Omega_4} 
\newcommand{\Tr}{{\rm Tr}\,}

\newcommand{\td}{\tilde d} 
\newcommand{\tc}{\tilde c} 

\def\Im{{\rm Im}\,}

\begin{document}

\title{Holographic QCD predictions for production and decay of pseudoscalar glueballs}


\author{Frederic Br\"unner}
\author{Anton Rebhan}
\affiliation{Institut f\"ur Theoretische Physik, Technische Universit\"at Wien,
        Wiedner Hauptstrasse 8-10, A-1040 Vienna, Austria}

\date{\today}

\begin{abstract}
The top-down holographic Witten-Sakai-Sugimoto model 
for low-energy QCD, augmented by finite quark masses, has recently been found
to be able to reproduce the decay pattern
of the scalar glueball candidate $f_0(1710)$ on a quantitative level.
In this Letter we show that this model
predicts a narrow pseudoscalar glueball heavier than the scalar glueball and
with a very restricted decay pattern involving $\eta$ or $\eta'$ mesons.
Production should be either in pairs or in
association with $\eta(')$ mesons. We discuss the prospect of
discovery in high-energy hadron collider experiments through
central exclusive production by comparing with $\eta'$ pair production.
\end{abstract}

\maketitle

\section{Introduction}

Quantum chromodynamics, the established theory of the strong interactions, predicts
\cite{Fritzsch:1972jv,*Fritzsch:1975tx,*Jaffe:1975fd}
the existence of flavor singlet mesons beyond those required by the quark model, because
in the absence of quarks gluons by themselves can form bound states.
However, the status of such ``glueball'' states in
the observed meson spectrum is still unclear and controversal
\cite{Bugg:2004xu,Klempt:2007cp,Crede:2008vw,Ochs:2013gi}.

In 1980, an isoscalar pseudoscalar with a mass of around 1.44 GeV which is copiously produced
in the gluon-rich radiative decays of $J/\psi$
was proposed as the first glueball candidate \cite{Donoghue:1980hw,*Ishikawa:1980xv,*Chanowitz:1980gu}.
Once named $\iota(1440)$ \cite{Edwards:1982nc}, this is now listed by the Particle Data Group \cite{PDG16}
as the two states $\eta(1405)$ and $\eta(1475)$. Together with $\eta(1295)$, this indeed would
give rise to a supernumerary state beyond the first radial excitations of the $\eta$ and $\eta'$ mesons,
with $\eta(1405)$ singled out as glueball candidate \cite{Masoni:2006rz}.

The situation thus appears to be analogous to the case of the scalar glueball,
which is widely considered to be responsible for a supernumerary state in the set of
isoscalar scalar resonances $f_0(1370)$, $f_0(1500)$, and $f_0(1710)$, where only
two are expected from the quark model (corresponding to $\bar{u}u+\bar{d}d$ and $\bar{s}s$).
Here the discussion is divided on the question which of the two heavier resonances
has the larger glueball contribution
\cite{Amsler:1995td,*Lee:1999kv,*Close:2001ga,*Amsler:2004ps,*Close:2005vf,*Giacosa:2005zt,*Albaladejo:2008qa,*Mathieu:2008me,*Janowski:2011gt,Janowski:2014ppa,Cheng:2015iaa,Close:2015rza,Frere:2015xxa}.

However, only the case of the scalar glueball candidates is supported by existing lattice QCD calculations
\cite{Morningstar:1999rf,Chen:2005mg} 
which consistently find that the lowest-lying glueball state has a mass of around 1.7 GeV and quantum numbers $J^{PC}=0^{++}$.
The lowest-lying pseudoscalar glueball state is instead found to have a mass of around 2.6 GeV,
somewhat higher than the $2^{++}$ tensor glueball with a mass of around 2.4 GeV.
Most lattice results have been obtained in the quenched approximation\footnote{It has been argued that the pseudoscalar sector may be particularly sensitive to unquenching
in Ref.~\cite{Gabadadze:1997zc}, but the estimated effects on the mass were of the order of 15\%,
whereas almost 50\% would be needed to bring the lattice result down to the mass of $\eta(1405)$.}, i.e.\
without dynamical quarks,
but recent unquenched lattice calculations \cite{Richards:2010ck,Gregory:2012hu,Sun:2017ipk}
have found no evidence for significant unquenching effects, which however should be expected if the pseudoscalar
glueball were to mix strongly with radially excited $\eta(')$ mesons. 
\1Moreover, Ref.~\cite{Sun:2017ipk} recently reported that correlation functions of
pseudoscalar gluonic operators built from Wilson loops did not show any trace of
the flavor singlet pseudoscalar meson 
states which can be found in the topological charge density correlator.
In fact, on the experimental side\0{} it is still
a controversial issue whether as many as three states $\eta(1295)$, $\eta(1405)$, and $\eta(1475)$
\1(and thus indication of the involvement of the pseudoscalar glueball
in this mass region)\0{}
really exist.\footnote{E.g., the existence of $\eta(1295)$ is questioned in \cite{Klempt:2007cp}, while
Ref.~\cite{Bugg:2009cf} came to the conclusion
that there is ``no evidence for two separate $\eta(1405)$ and $\eta(1475)$
from the present data'' and only one $\eta(1440)$ is actually required.}

We therefore assume that \1(contrary to the models used in Ref.~\cite{Cheng:2008ss,Ambrosino:2009sc,Gutsche:2009jh})\0{}
the pseudoscalar glueball \1does not make its appearance in the known $\eta$ mesons in the 1400 MeV region, but that it\0{}
still has to be discovered and that it should be searched for in the mass range 2--3 GeV.
Unfortunately, lattice QCD does not (yet) give information on the production and decay
patterns of a pseudoscalar glueball, whereas phenomenological models are weakly constrained
with regard to the particular form of pseudoscalar glueball interactions.\footnote{In Ref.~\cite{Eshraim:2012jv}
a unique form of the interaction Lagrangian for extended linear sigma models has been posited, 
where only the coupling strength
is left undetermined, but in a subsequent extension \cite{Eshraim:2016mds} more possibilities
were introduced.}

In this work we show that rather specific predictions can be obtained from the 
Witten-Sakai-Sugimoto (WSS) model for low-energy QCD, which is a top-down
string-theoretic construction in the large color number ($N_c$) limit
with only one free dimensionless parameter.
Extrapolated to $N_c=3$, it 
reproduces several experimental results in hadron physics to within 10-30\% \cite{Rebhan:2014rxa,Brunner:2015oqa}.
In Ref.~\cite{Brunner:2015oqa}
we have applied this model to calculate decay rates of scalar and tensor
glueballs in the chiral limit, and in \cite{1504.05815,1510.07605} with quark masses
included. In the latter case we found 
a strong ``nonchiral enhancement'' of the decay of a
predominantly dilatonic glueball into kaons and $\eta$ mesons
which quantitatively agrees remarkably well with the data for the glueball candidate $f_0(1710)$
as far as presently known (provided the not-yet-measured decay rate into $\eta\eta'$ pairs
is sufficiently small \cite{1510.07605}). This suggests that $f_0(1710)$
could be a nearly pure glueball, in agreement
with recent phenomenological models that favor $f_0(1710)$ as the scalar
glueball \cite{Janowski:2014ppa,Cheng:2015iaa} with comparatively
small admixture of light quarkonia.

While in Ref.~\cite{Brunner:2015oqa} our WSS model 
prediction for the width of the tensor glueball
of mass $\gtrsim2$ GeV was very large, perhaps too large to be clearly observable,
here we arrive at the prediction of a narrow pseudoscalar glueball state with a very restricted decay pattern,
which will be a conspicuous feature as long as mixing with quarkonia is small.
The specific interactions also suggest that the pseudoscalar glueball may be difficult to
produce in radiative charmonium decay, but
could be a very interesting object for glueball searches in central exclusive production (CEP) experiments
at sufficiently high energies.

\section{Effective Lagrangian for pseudoscalar glueball interactions}

The WSS model \cite{Witten:1998zw,Sakai:2004cn,Sakai:2005yt}
is a gauge/gravity-dual model for 
nonsupersymmetric low-energy QCD based on D4 branes in
type-IIA supergravity compactified on a circle and subjected to
a consistent truncation of 
Kaluza-Klein states, with $N_f\ll N_c$ chiral quarks added through probe D8 branes. It
possesses an interesting spectrum 
of glueball states 
with $J^{PC}=0^{++},2^{++},0^{-+},1^{+-},1^{--}$ \cite{Brower:2000rp}
whose mass scale is set by the Kaluza-Klein mass $\MKK$. 
The resulting effective theory involves
Goldstone pseudoscalars for nonabelian chiral symmetry breaking
and a tower of vector and axial vector mesons. 

Fixing $\MKK$ through the experimental value of the $\rho$ meson mass and varying 
the 't Hooft coupling $\lambda=16.63\ldots12.55$ such that either the pion decay constant
or the string tension in large-$N_c$ lattice simulations \cite{Bali:2013kia} is matched
leads to quantitative predictions which are in the right ballpark when extrapolated to $N_c=3$ QCD,
\1including a value for the gluon condensate 
\be\label{gluoncondensate}
C^4\equiv\left<\frac{\alpha_s}{\pi}G_{\mu\nu}^a
G^{a\mu\nu}\right>=\frac1{2\pi^2}\left<\Tr F^2\right>=
\frac{4N_c}{3^7\pi^4}\lambda^2\MKK^4
\ee
that is close to that obtained
by SVZ sum rules \cite{Rebhan:2014rxa}.
Moreover\0, it reproduces remarkably well the observed
hadronic decay rates of
the $\rho$ and the $\omega$ mesons, which motivates the use of the WSS model also as
a model for glueball decay \cite{Hashimoto:2007ze,Brunner:2015oqa}.
In Ref.~\cite{Brunner:2015oqa} we argued, however, that the lightest scalar glueball
mode considered in Ref.~\cite{Hashimoto:2007ze} which comes from an ``exotic polarization''
of the dual graviton along the compactified direction (denoted by $G_E$ in the following)
should be discarded and that
instead the predominantly dilatonic mode ($G_D$) be identified with the glueball ground state.

The WSS model correctly incorporates the nonabelian chiral anomaly \1of QCD and the resulting
Wess-Zumino-Witten term as well as the U(1)$_A$ anomaly\0{} and
the Witten-Veneziano mechanism for giving
mass to the flavor singlet pseudoscalar $\eta_0$ with \cite{Armoni:2004dc,Barbon:2004dq,Sakai:2004cn} 
\be\label{mWV2}
m_{0}^2=\frac{N_f}{27\pi^2 N_c}\lambda^2\MKK^2,
\ee
leading to $m_{0}=730$ - $967$ MeV for $\lambda$ between 12.55 and 16.63.
Introducing explicit quark mass terms in the effective Lagrangian such that
physical pion and kaon masses are matched leads to $\eta$ and $\eta'$ masses
that agree with real QCD to within $\lesssim10\%$
\cite{1504.05815,1510.07605}. As mentioned above,
the flavor-asymmetric decay pattern observed for the scalar glueball candidate $f_0(1710)$
can be reproduced quantitatively with $G_D$, if the
(as yet undetermined) parameter for scalar glueball couplings to explicit quark mass terms
is chosen such that the rate of decay into mixed $\eta\eta'$ pairs
remains small. 

The interaction Lagrangian of the pseudoscalar glueballs is the same for both,
the chiral and the massive version of the WSS model. The pseudoscalar glueball modes
are provided by a Ramond-Ramond (RR) 1-form field $C_1$ which plays the central role in producing
the  Witten-Veneziano mass $m_0$. 
Following the notation of Ref.~\cite{Sakai:2004cn},
the action for $C_1$ is given by 
\be\label{SC1}
S_{C_1}=-\frac1{4\pi(2\pi l_s)^6}\int d^{10}x \sqrt{-g} |\tilde F_2|^2.
\ee
As reviewed in the Appendix, anomaly cancellation requires that 
$\tilde F_2$ is a gauge invariant combination of $F_2=dC_1$ and
the field 
\be
\eta_0(x)=\frac{f_\pi}{\sqrt{2N_f}}\int dz \,{\rm Tr} A_z(z,x)
\ee
with $z$ parametrizing the radial extent of the joined
D8 and anti-D8 branes on which the flavor gauge field $A$ lives.

Inserting a mode expansion of the RR 1-form field $C_1$ with
4-dimensional pseudoscalar glueball fields $\tilde{G}^{(n)}(x)$, $n=1,\ldots$,
together with scalar and tensor glueball fields entering through
the metric in $S_{C_1}$ 
leads to the effective 4-dimensional Lagrangian 
\bea\label{LeffC1}
\mathcal L_{C_1}^{\rm eff}&=&-\frac12 \partial_\mu \tilde{G}\,\partial^\mu \tilde{G}-\frac12 m_P^2 \tilde{G}^2
-\frac12 m_0^2 \eta_0^2\nn\\
&+&\mathcal L_{\eta_0^2 G}+\mathcal L_{\tilde{G}\eta_0 G}+\mathcal L_{\tilde{G}^2 G}+O(G^2_{D,E,T})
\eea
(suppressing the summation over the mode number index $(n)$).
Here $O(G^2_{D,E,T})$ denotes higher-order interactions involving terms 
quadratic in $\tilde{G},\eta_0$ and quadratic or higher in the glueball fields arising
from metric fluctuations
(the tensor glueball field $T^{\mu\nu}$ appears at most linearly, but also has
interactions involving arbitrarily high powers of the scalar glueball field).

The mass of the lowest pseudoscalar glueball mode ($n=1$) is \cite{Brower:2000rp}
$M_P\approx 1.885\MKK$, which like in lattice QCD results is above the mass
of the scalar and tensor glueballs with $M_{D}=M_T\approx 1.567\MKK$.
With $\MKK=949$ MeV from having matched the mass of the $\rho$ meson, 
$M_D\approx 1487$ MeV and $M_P\approx 1789$ MeV, but in the eventual
applications we shall leave $M_P$ a free parameter and either keep
$M_D$ at 1.5 GeV which approximately matches the mass of $f_0(1500)$ 
or artificially raise its mass to the mass of $f_0(1710)$.

Note that Eq.~(\ref{LeffC1}) contains a mass term for the flavor singlet $\eta_0$
\cite{Sakai:2004cn}, but no
mixing of the pseudoscalar glueball modes $\tilde{G}^{(n)}$ with $\eta_0$.\footnote{This
feature is due the fact the WSS model corresponds to QCD in the 't Hooft limit
$N_c\gg 1$ but $N_f\sim 1$. In the bottom-up holographic model of Ref.~\cite{Arean:2016hcs}, where
the Veneziano limit$N_c\gg 1$ and $N_f/N_c\sim 1$ is taken, mixing of
pseudoscalar glueballs and $\eta_0$ appears at leading order, but in a way
that depends strongly on the choice of potentials.}
\1As shown in the Appendix,\0{}
terms proportional to $\eta_0\tilde{G}^{(n)}$ vanish in the unperturbed background
geometry, but arise in the presence of metric fluctuations dual to scalar glueballs.
In the WSS model, such terms are the only ones which can mediate a decay
of pseudoscalar glueballs. Explicitly they read (keeping the exotic glueball
mode $G_E$ for completeness)
\bea\label{LGTeG}
\mathcal L_{\tilde{G}\eta_0 G}&=&
\td_0\, \tilde{G}\, \eta_0\, G_D+ \tc_0\, \tilde{G}\, \eta_0\, G_E\nn\\
&&+ \frac{\tc_0'}{M_E^2}\, \partial_\mu \tilde{G}\, \eta_0\, \partial^\mu G_E 
+\tc_0''\, \tilde{G}\, \eta_0\, \frac{\Box-M_E^2}{M_E^2} G_E\quad
\eea
with the numerical results for the coupling constants for
the lowest pseudoscalar glueball mode listed in Table \ref{tabcGt}
(their integral representations will be given elsewhere).

The part of the action which leads to the Witten-Veneziano mass term 
also gives rise to interactions with scalar glueballs which were obtained 
(on-shell) in \cite{1504.05815}. 
To linear order in glueball fields the corresponding
interaction Lagrangian reads (also including an extra off-shell contribution
for the exotic mode $G_E$)
\bea\label{LeeG}
\mathcal L_{\eta_0^2 G}
&=&\frac12 m_0^2 \eta_0^2\left(3 d_0 G_D-5\breve c_0 G_E\right)\nn\\
&&+\frac12 \bar c_0 m_0^2 \eta_0^2 \frac{\Box-M_E^2}{M_E^2} G_E. 
\eea
There are also interaction terms of the form $(\partial\eta_0)^2G_{D,E,T}$
coming from the DBI action of the D8 branes, which can be found in Ref.~\cite{Brunner:2015oqa},
as well as natural-parity violating terms $\eta_0 G_T^2$ from Chern-Simons action
of the D8 branes, which have been obtained in Ref.~\cite{Anderson:2014jia}.

\begin{table}
\begin{tabular}{c|r}
\toprule
coeff.&value\\
 \colrule
$\bar d_0$&$17.915\,\lambda^{-1/2} N_c^{-1}\MKK^{-1}$\\
$\td_0$&$2.5833\,\lambda^{1/2} N_f^{1/2} N_c^{-3/2}\MKK$\\
$\td_1$&$42.484\,\lambda^{-1/2} N_c^{-1}\MKK^{-1}$\\
$\td_2$&$27.106\,\lambda^{-1/2} N_c^{-1}\MKK^{-1}$\\
\colrule
$\breve c_0$&$15.829\,\lambda^{-1/2} N_c^{-1}\MKK^{-1}$\\
$\bar c_0$&$26.837\,\lambda^{-1/2} N_c^{-1}\MKK^{-1}$\\
$\tc_0$&$-4.8795\,\lambda^{1/2} N_f^{1/2} N_c^{-3/2}\MKK$\\
$\tc'_0$&$1.6306\,\lambda^{1/2} N_f^{1/2} N_c^{-3/2}\MKK$\\
$\tc''_0$&$2.0502\,\lambda^{1/2} N_f^{1/2} N_c^{-3/2}\MKK$\\
\botrule
\end{tabular}
\caption{Coupling constants in the glueball interaction Lagrangians (\ref{LGTeG}), (\ref{LeeG}), and (\ref{LGT2G}).}
\label{tabcGt}
\end{table}

Interaction terms involving pairs of pseudoscalar glueballs and a scalar or tensor glueball
are given by
\bea\label{LGT2G}
&&\mathcal L_{\tilde{G}^2 G}
=\td_1 \biggl[ \frac12 \partial_\mu \tilde{G}\, \partial^\mu \tilde{G} 
-\frac18 \partial_\mu \tilde{G} \, \partial_\nu \tilde{G} \frac{\partial^\mu\partial^\nu}{\Box} \biggr] G_D\nn\\
&&+\frac12 \td_2 m_P^2 \tilde{G}^2 G_D +\frac{\sqrt{6}}{8} \td_1 \partial_\mu \tilde{G}\,\partial_\nu \tilde{G}\, T^{\mu\nu}
+\mathcal L_{\tilde{G}^2 G_E}.
\eea
(The more unwieldy expression $\mathcal{L}_{\tilde{G}^2G_E}$
will be given elsewhere.)

\section{Decay pattern of the pseudoscalar glueball}

The only interaction terms arising within 
the WSS model that are relevant for the decay of pseudoscalar glueballs are contained in (\ref{LGTeG}).
They differ strongly from the leading interaction
terms that have been assumed previously in phenomenological models.


Rosenzweig et al.\ \cite{Rosenzweig:1981cu,Rosenzweig:1982cb} have assumed that
the chiral anomaly is not saturated by $\eta_0$ alone, but involves a further physical pseudoscalar field 
($\tilde{G}_2$),\footnote{In Ref.~\cite{Rosenzweig:1981cu} $\tilde{G}_1$ is an auxiliary field with wrong-sign mass term
that can be replaced by $\Im\log\det\Sigma$, which is essentially $\eta_0$, through its
algebraic equations of motion.}
which couples to the imaginary part of $\log\det\Sigma$, where $\Sigma$ is the matrix of
$q \bar q$ states (which is unitary in the nonlinear sigma model, involving only the
pseudoscalars, but unrestricted in linear sigma models \cite{Rosenzweig:1979ay} so that it also accommodates scalar mesons).
While a natural possibility \cite{Rosenzweig:1981cu} would be to identify $\tilde{G}_2$ with the radial excitation of $\eta_0$,
it was proposed to identify $\tilde{G}_2$ with the pseudoscalar glueball instead. Originally used in the
context of the glueball candidate $\iota(1440)$, this approach was also adopted in the extended linear
sigma model of Ref.~\cite{Eshraim:2012jv} for pseudoscalar glueballs with a mass suggested
by lattice QCD. The dominant decay mode of a pseudoscalar glueball
in this approach turns out to be $K\bar K\pi$ (branching ratio $\mathcal{B}\approx 1/2$) 
followed by $\eta\pi\pi$ ($\mathcal{B}\approx 1/6$)
and $\eta'\pi\pi$ ($\mathcal{B}\approx 1/10$).

Using large-$N_c$ chiral Lagrangians, Gounaris et al.~\cite{Gounaris:1988rp}
argued that there should be no coupling of the pseudoscalar glueball to $\Im\log\det\Sigma$.
Instead, a coupling to $\Im {\rm tr}\,\mathcal M_q \Sigma$ was considered so that the pseudoscalar
glueball is stable in the limit of massless quarks ($\mathcal M_q$ being the quark mass matrix).
This again gives a dominant decay mode $K\bar K\pi$, but with $\eta\pi\pi$ being more strongly suppressed
(parametrically by a factor $m_{\pi}^2/m_K^2$).

In agreement with the considerations of Ref.~\cite{Gounaris:1988rp}, the WSS model,
which also corresponds to a large-$N_c$ chiral Lagrangian, does not lead to
a coupling of the pseudoscalar glueball to $\Im\log\det\Sigma$. However, its extension to finite
quark masses (either through world-sheet instantons \cite{Aharony:2008an,*Hashimoto:2008sr,*McNees:2008km} 
or open-string tachyon condensation \cite{0708.2839,*Dhar:2007bz,*Dhar:2008um,*Niarchos:2010ki})
does not naturally lead to a coupling to $\Im{\rm tr}\,\mathcal{M}_q\Sigma$, because
Ramond-Ramond fields do not couple directly to fundamental strings. In the WSS model,
the only coupling linear in $\tilde{G}$ is to $\eta_0G$. 
This suggests that the pseudoscalar glueball should decay dominantly in $\eta(')$ and
the $f_0$ meson which corresponds to the scalar glueball, or $\eta(')$ and decay products
of the latter. According to the WSS model, the decay mode $K\bar K\pi$ 
that is obtained as the dominant one in the approaches
mentioned above should instead be strongly suppressed.


\begin{figure}
\includegraphics[width=0.475\textwidth]{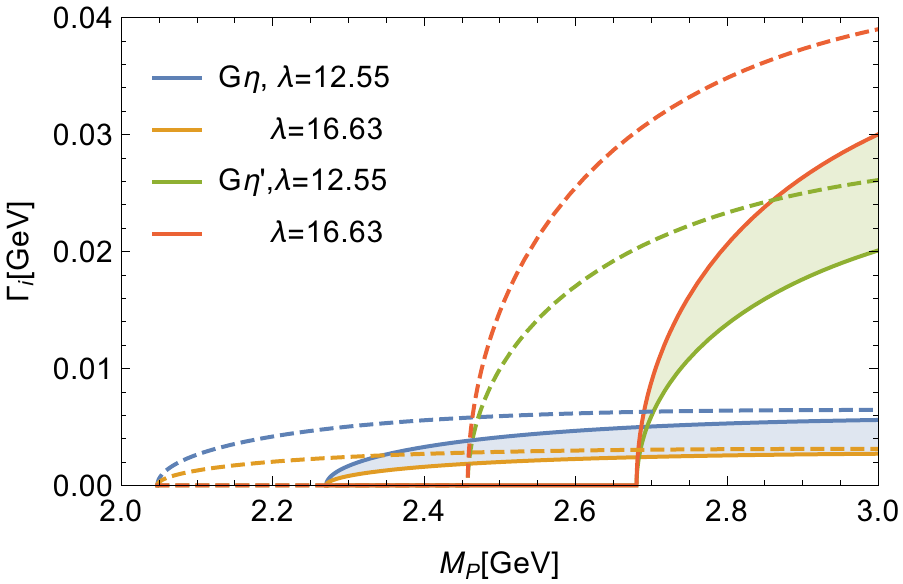}
\caption{Partial width of resonant decay $\tilde{G}\to G\eta{(')}$ (neglecting finite width of scalar glueball)
for a predominantly dilatonic scalar glueball $G_D$ with
mass $m_D=1.5$ GeV (dashed lines) and 1.723 GeV (full lines), the latter
corresponding to $f_0(1710)$. 
}
\label{figdecayGTetaG}
\end{figure}

\begin{figure}
\includegraphics[width=0.475\textwidth]{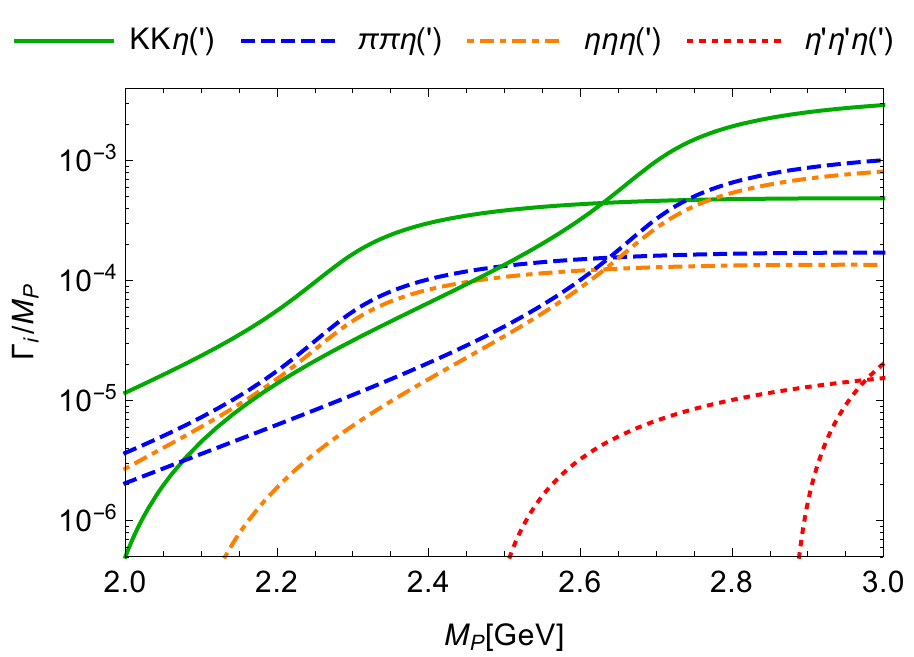}
\caption{Partial widths of resonant and non-resonant decays $\tilde{G}\to G\eta{(')}\to PP\eta{(')}$
where $P=K,\pi,\eta,\eta'$ assuming the decay pattern for the scalar glueball $G=f_0(1710)$ obtained
in Ref.~\cite{1504.05815}. (The two cases $PP\eta$ and $PP\eta'$ are plotted in the same style but can be distinguished easily by the later onset of $PP\eta'$ which dominates at sufficiently high values of $M_P$.)}
\label{figdecayGTetaPP}
\end{figure}

When the mass of the pseudoscalar glueball is larger than the mass of the scalar glueball plus the $\eta(')$ mass,
the scalar glueball can be produced on-shell. The resulting decay width is displayed in Fig.~\ref{figdecayGTetaG}
as a function of the pseudoscalar glueball mass for the glueball mode $G_D$ with mass 1.5 GeV and also when
raised in mass to match $f_0(1710)$, which in Ref.~\cite{1504.05815} we found 
to be favored by the WSS model.\footnote{\1As in Refs.~\cite{Brunner:2015oqa,1504.05815,1510.07605} we discard the ``exotic'' scalar glueball
mode $G_E$, assuming it has no counterpart in QCD.
If we had identified the 
mode $G_E$ with the lowest scalar glueball\0{}
and raised its mass (which is originally only 855 MeV) to the mass of $f_0(1500)$ or $f_0(1710)$,
Fig.~\ref{figdecayGTetaG} would look very similar, but the decay width would be about a factor of 10 larger.} For the latter case,
Fig.~\ref{figdecayGTetaPP} shows the (not necessarily resonant) dimensionless partial decay widths $\Gamma_i/M_P$
for $\tilde{G}\to G\eta{(')}\to PP\eta{(')}$
where $P=K,\pi,\eta,\eta'$ with the decay pattern for the scalar glueball $G=f_0(1710)$ obtained
in Ref.~\cite{1504.05815}. 
With $m_P\sim 2.6$ GeV as predicted by lattice QCD, the pseudoscalar glueball is predicted to be a rather narrow
state; for $m_P\lesssim2.3$ GeV it would be extremely narrow \1(in this case
it is of course probable that other, subleading decay channels which are beyond
the WSS model become equally if not more important)\0.

\section{Production of pseudoscalar glueballs}

While scalar and tensor glueballs couple directly to $q\bar{q}$ mesons, pseudoscalar glueballs
do so only through the former in the WSS model, \1because the $C_1$ Ramond-Ramond field does not couple
directly to the DBI and CS action of flavor D8 branes\0. This suggests that pseudoscalar glueballs are not as easily formed in radiative decays of $J/\psi$
as the other glueballs, but they would have to arise from excited scalar or tensor glueballs
decaying into $\eta(')\tilde{G}$ or $\tilde{G}\tilde{G}$ pairs. The thresholds for these processes
are thus above the mass of the $J/\psi$ so that excited $\psi$ mesons or $\Upsilon$ would be required.
\1Creation of $\eta(')\tilde{G}$ or $\tilde{G}\tilde{G}$ pairs via virtual
scalar and tensor glueballs would also be a possibility for the planned glueball searches
in proton-antiproton collisions in the PANDA experiment \cite{Lutz:2009ff} at FAIR.\0%
\footnote{\1In Ref.~\cite{Eshraim:2012rb} a chirally invariant coupling of
the pseudoscalar glueball to nucleons and their chiral partners in the so-called mirror assignment
was considered. In the WSS model, baryons are described by Skyrmion-like solitons of the effective action of the flavor branes, which likewise excludes a direct coupling to the pseudoscalar glueball
at the same order as the direct couplings to scalar and tensor glueballs.}

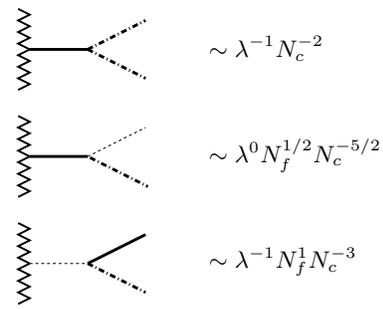
\begin{figure}
\centerline{\hfil\includegraphics[width=0.1\textwidth]{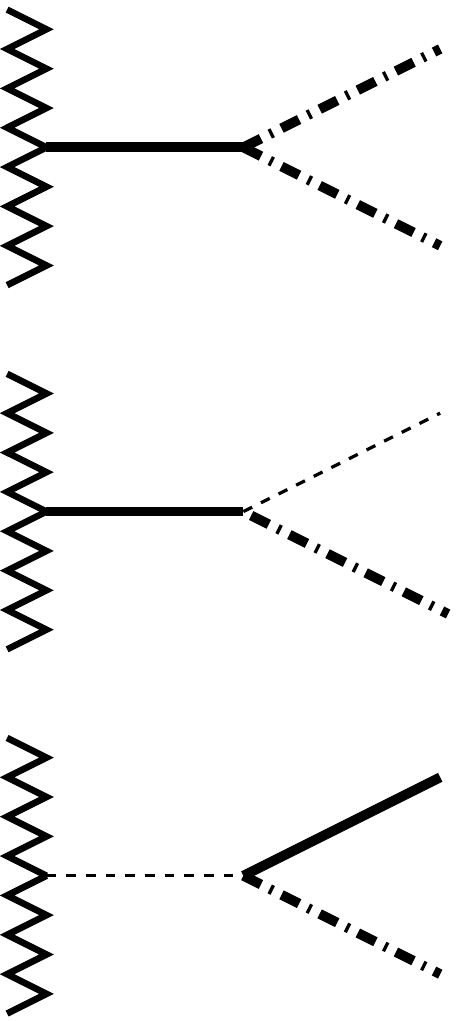}\hfil\hfil\hfil}
\begin{picture}(0,0)
\put(15,105){$\sim\lambda^{-1}N_c^{-2}$}
\put(15,65){$\sim\lambda^{0}N_f^{1/2}N_c^{-5/2}$}
\put(15,25){$\sim\lambda^{-1}N_f^{1}N_c^{-3}$}
\end{picture}
\caption{Parametric orders of the production amplitudes of pseudoscalar glueballs 
($\tilde G\tilde G$, $\eta(')\tilde G$, and $G\tilde G$, respectively)
in double Pomeron
or double Reggeon exchange. (Dotted, full, and dash-dotted lines represent
$\eta(')$, $G$, and $\tilde G$, respectively. In the uppermost diagram the full line
stands for $G$ or $G_T$.)}
\label{figCEP}
\end{figure}

Another possibility is central exclusive production (CEP) in high-energy hadron collisions
through double Pomeron or Reggeon exchange (corresponding
to $G_T$ and $(\rho,\omega)$ trajectories; pion and scalar glueball exchanges are subdominant
at high energies). 
The parametric orders of the corresponding amplitudes are
shown in Fig.~\ref{figCEP}.
Production of $\tilde G\eta_0$ occurs only via virtual scalar glueballs, whereas
production of $\tilde G\tilde G$ can additionally proceed through virtual tensor glueballs. 
Also shown is the possibility of $G\tilde G$ production through the natural-parity violating
coupling of $\eta_0$ to two tensor glueballs (Pomerons), which is provided by the
Chern-Simons part of the action of the D8 branes and which was recently studied
within the WSS model in Ref.~\cite{Anderson:2014jia}.\footnote{A natural-parity violating
coupling of $\eta_0$ also exists with Reggeons. Fig.~\ref{figCEP} gives the parametric order for double Pomeron
exchange, which is down by a factor $1/N_c$ compared to Reggeons, but becomes
dominant at sufficiently high energies.}

Associated production of pseudoscalar glueballs with either $\eta(')$ or other glueballs
is presumably beyond the reach of the older fixed-target experiments searching for glueballs,
but seem to be an exciting possibility for the new generation of CEP experiments
at the LHC.

\begin{figure}[t]
\includegraphics[width=0.475\textwidth]{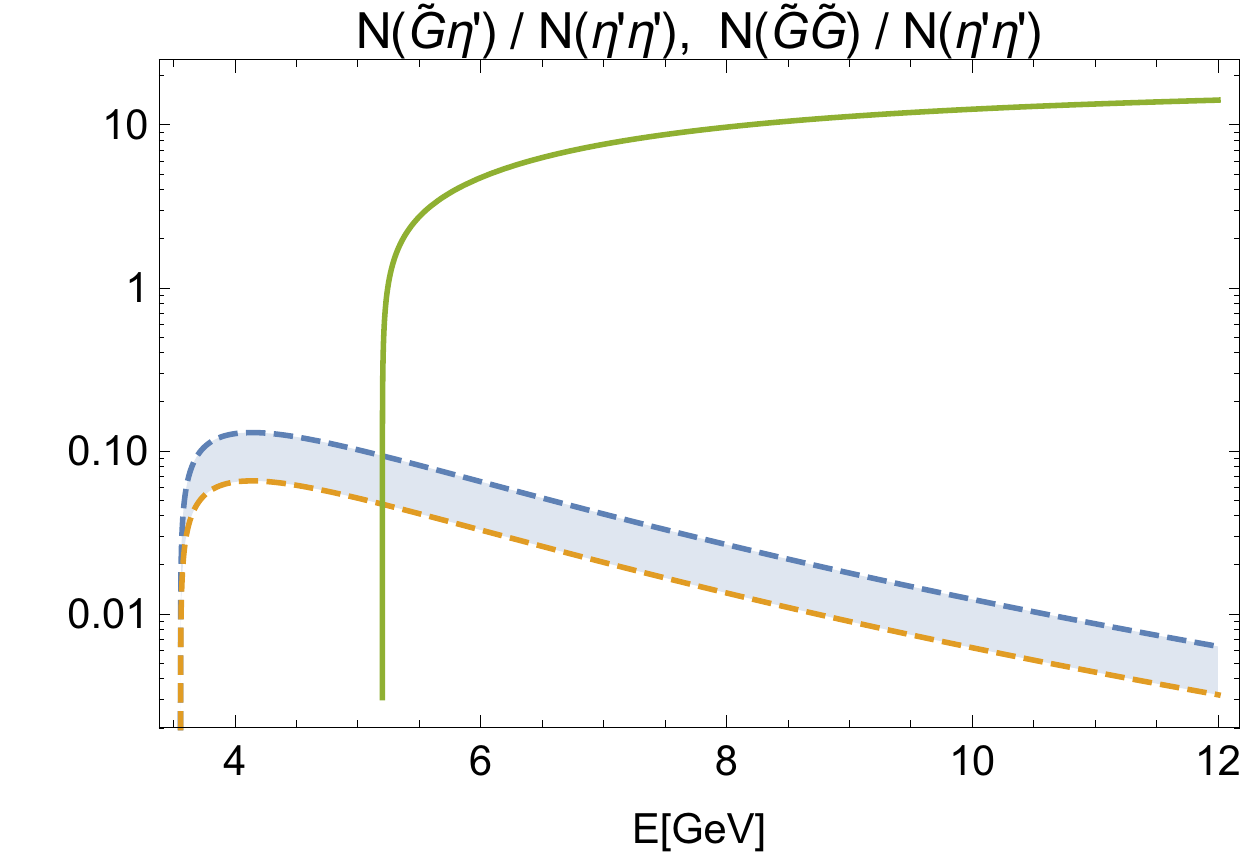}
\caption{Production of $\tilde{G}\tilde{G}$ and $\tilde{G}\eta'$ pairs versus $\eta'\eta'$ from a virtual scalar glueball $G_D$ for
a pseudoscalar glueball mass of 2.6 GeV as functions of the center of mass energy of
the produced pair. The full line gives 
the ratio of the numbers of produced pairs
$N(\tilde{G}\tilde{G})/N(\eta'\eta')$,
which is independent of the 't Hooft coupling; upper and lower dashed lines
correspond to $N(\tilde{G}\eta')/N(\eta'\eta')$ with 't Hooft coupling 12.55 and 16.63, respectively.}
\label{fig:pscgbproduction}
\end{figure}

Calculation of the corresponding production cross sections within the WSS model
could be attempted by employing the techniques used in Ref.~\cite{Anderson:2014jia} for
$\eta$ and $\eta'$ production \1(see also \cite{Herzog:2008mu,Brower:2012mk})\0, 
but will be left for future work.
In this Letter we only present results for the ratio of production rates of
$\tilde{G}\eta'$ and $\tilde{G}\tilde{G}$ pairs over $\eta'\eta'$
pairs,\footnote{The production rate of $\tilde{G}\eta$ has a smaller
threshold and thus larger phase space but is reduced
by a factor $(\tan\theta_P)^2\sim$ 0.1.} when both are produced through a virtual $G_D$ glueball.
This ratio is fixed by the vertices obtained above together with the results
obtained in Ref.~\cite{1504.05815}, and the result is shown in 
Fig.~\ref{fig:pscgbproduction} for the range of 't Hooft coupling discussed above.
The amplitude $\mathcal{M}(G^*\to\tilde{G}^2)\sim\lambda^{-1/2}N_c^{-1}$ 
is parametrically 
of the same
order as $\mathcal{M}(G^*\to\eta'^2)$ so that
the ratio $N(\tilde{G}\tilde{G})/N(\eta'\eta')$ is particularly well determined (at least for fixed meson masses
in the scenario of Ref.~\cite{1504.05815}).
The results in Fig.~\ref{fig:pscgbproduction} indicate that 
CEP of $\eta'\tilde{G}$ is only one order of magnitude below CEP of $\eta'\eta'$, while
above the threshold for $\tilde{G}\tilde{G}$ pairs, production of the latter is even up to one order of magnitude larger
than CEP of $\eta'\eta'$.

Central exclusive production of $\eta'$ pairs has been studied
in the Durham model in Ref.~\cite{Harland-Lang:2013ncy}, where its production cross section was estimated.
For example, at $\sqrt{s}=1.96$~TeV this work obtained 
$\sigma(\eta'\eta')/\sigma(\pi^0\pi^0)\sim 10^3\ldots10^5$
assuming sufficiently high transverse momentum such that a perturbative approach becomes justified.

Since small transverse momentum is expected to provide a glueball filter \cite{Close:1997pj,Kirk:2014nwa}
and the production of $\tilde{G}$ together with another $\tilde{G}$ or $\eta(')$ according to the present model
proceeds through virtual scalar glueballs,
the kinematical regime of small transverse momentum (small azimuthal angle
$\phi_{pp}$) would be particularly interesting for the search of
pseudoscalar glueballs.

To summarize, the WSS model, which is based on the 't Hooft limit of large $N_c\gg N_f$
where mixing of glueballs with $q\bar{q}$ states is suppressed,
suggests a very restricted decay pattern of a rather narrow
pseudoscalar glueball $\tilde{G}$, namely decay into an $\eta{'}$ meson together with
a scalar glueball, with the latter decaying mostly into pairs of pseudoscalar mesons.
\1In particular, the $K\bar K\pi$ decay mode obtained in many other models
is found to be suppressed, because the WSS model does not directly couple the pseudoscalar glueball
mode carried by the Ramond-Ramond gauge field $C_1$
to fundamental strings and flavor branes. This is certainly not a universal feature of
holographic models and thus will not necessarily hold in other (e.g., bottom-up) holographic
approaches to QCD, but the (top-down) WSS model appears to be particularly attractive because
it incorporates nonabelian chiral symmetry breaking as well as the anomaly structure of QCD
in a most natural way.

By the same token, the WSS model predicts the production of pseudoscalar glueballs\0{}
to proceed through excited scalar or tensor glueballs
decaying into $\eta{'}\tilde{G}$ or $\tilde{G}\tilde{G}$ pairs so that the threshold
is above radiative $J/\psi$ decays. This could explain why no pseudoscalar glueball
candidate with mass in the range of lattice predictions
has as yet been found there. Instead, searches in excited charmonium or $\Upsilon$ decays and
CEP experiments at high-energy hadron colliders 
\1as well as proton-antiproton collisions at FAIR
\0{}
should have the potential for finally
discovering the pseudoscalar glueball, with production cross-sections comparable
to those of $\eta'\eta'$ pairs.

\begin{acknowledgments}
We thank Claude Amsler, Paolo Gandini, Nelia Mann, Denis Parganlija, and Ulrich Wiedner for discussions and correspondence.
This work was supported by the Austrian Science
Fund FWF, project no. P26366, and the FWF doctoral program
Particles \& Interactions, project no. W1252.
\end{acknowledgments}

\appendix

\section{Mode expansion of the $C_1$ Ramond-Ramond field sector} 

In this appendix we recapitulate 
some fundamental properties of the WSS model, in particular concerning
the sector involving the $C_1$ Ramond-Ramond field, together with the mode expansion of the latter
that is needed to study pseudoscalar glueball interactions.

The metric in the WSS model reads 
\bea\label{ds210}
ds^2&=&\left(\frac{u}{\RSS}\right)^{3/2} \left[\eta_{\mu\nu}dx^\mu dx^\nu
+f(u)(dx^4)^2\right]\nonumber\\
&&+\left(\frac{\RSS}{u}\right)^{3/2}\left[\frac{du^2}{f(u)}+u^2 d\Omega_4^2 \right]
\eea
with $f(u)=1-(\uKK/u)^3$; the nonconstant dilaton is given by 
\be\label{Phibackground}
e^\Phi=(u/\RSS)^{3/4}.
\ee

The parameters of the dual field theory 
are given by \cite{Kruczenski:2003uq,Sakai:2004cn,Sakai:2005yt,Kanitscheider:2008kd}
\bea\label{gYMNc}
\gYM^2=\frac{g_5^2}{2\pi R_4}=2\pi\gs\ls\MKK,\quad
\RSS^3=\pi \gs \Nc \ls^3.\qquad
\eea

The RR 1-form field $C_1=C_\tau(u,x) d\tau$ contains pseudoscalar glueball modes.
For nonvanishing $\theta$-parameter, it also carries nonvanishing flux through the 2-plane
parametrized alternatively by $(u,\tau)$ or $(y,z)$ with $y=0$ being the position of the stack of D8 branes.
Anomaly cancellation requires that $C_1$ transforms nontrivially under U(1) flavor gauge field transformations.
This can be taken into account by replacing
its field strength $F_2=dC_1$ in the 10-dimensional action for $C_1$ by
the gauge invariant combination
\bea
\tilde F_2&=&dC_1+{\rm tr}(A)\wedge \delta(y)dy\\
&=&dC_1'+\frac{c}{u^4}
\left( \theta+\frac{\sqrt{2N_f}}{f_\pi}\eta_0(x) \right)du\wedge d\tau\nonumber
\eea
where $C_1'$ is a reduced RR 1-form field with zero net flux through the $(u,\tau)$-plane and
\be
c=\frac{3 \uKK^3}{\delta\tau},\quad \delta\tau\equiv{2\pi/\MKK}
\ee
such that in the absence of $C_1'$
\be
\partial_u(\sqrt{-g}g^{uu}g^{\tau\tau} \tilde F_{u\tau})=\partial_u(u^4 \tilde F_{u\tau})=0
\ee
and
\be
\int \tilde F_{u\tau}du\wedge d\tau
=
\left( \theta+\frac{\sqrt{2N_f}}{f_\pi}\eta_0(x) \right) ,
\ee
\1with $f_\pi^2=\lambda N_c \MKK^2/(54\pi^4)$.\0{}
(Since we shall be setting $\theta=0$ in the end, we are ignoring here
that a finite $\theta$ leads to backreactions on the metric, which
have been worked out in \cite{Bigazzi:2015bna}. A priori, terms involving higher powers of $\theta$
require also contributions with higher powers of $\eta_0$ fields.
We have checked, however, that for $\theta=0$
inclusion of this backreaction does not lead to additional vertices involving
$\eta_0$ and pseudoscalar glueball modes beyond those worked out below.)

The action for the 1-form RR field is given by
\be
S_{C_1}=-\frac1{4\pi(2\pi l_s)^6}\int d^{10}x \sqrt{-g} |\tilde F_2|^2.
\ee

The reduced $C_1'$ will be expanded in pseudoscalar glueball modes, $C_1'=C'_\tau d\tau$ and
\be
C'_{\tau}(u,x)=\sum_{n=1}^\infty V^{(n)}(\bar u)\tilde G^{(n)}(x)
\ee
with radial mode functions 
$V^{(n)}(\bar u)=f(\bar u)\bar V^{(n)}(\bar u)$
satisfying
\be
-\bar u^{-1}\frac{d}{d\bar u}\left(\bar u^4 \frac{d}{d\bar u} 
\left[f(\bar u) \bar V^{(n)}(\bar u)\right] \right)=\frac94 \frac{(M_P^{(n)})^2}{\MKK^2}\bar V^{(n)}(\bar u)
\ee
where $\bar u=u/\uKK$ and $f(\bar u)=1-\bar u^{-3}$. 
The two lowest normalizable solutions with $V^{(n)}(\bar u=1)=V^{(n)}(\infty)=0$ but $\bar V^{(n)}(\bar u=1)\not=0$
have the eigenvalues $M_P^{(1)}\approx 1.885\MKK$ and $M_P^{(2)}\approx 2.838\MKK$, respectively.

With this mode expansion which keeps all fields independent of the compactified coordinate $\tau$ and
the coordinates of the $S^4$ we have
\bea\label{SC1-10}
S_{C_1}&=&-\frac1{4\pi(2\pi l_s)^6}\int d^{10}x \sqrt{-g}\biggl\{
g^{mn}g^{\tau\tau}\6_m C'_\tau \6_n C'_\tau\nn\\
&&+g^{uu}g^{\tau\tau}\frac{c^2}{u^8}\left( \theta+\frac{\sqrt{2N_f}}{f_\pi}\eta_0(x) \right)^2\nn\\
&&+2g^{mu}g^{\tau\tau}\6_m C'_\tau \frac{c}{u^4}\left( \theta+\frac{\sqrt{2N_f}}{f_\pi}\eta_0(x) \right)\biggr\}
,\quad
\eea
with indices $m,n\in\{0,1,2,3,u\}$.

\1
Inserting the background metric of the WSS model and setting $\theta=0$ produces the kinetic terms
in (\ref{LeffC1}) with Witten-Veneziano mass (\ref{mWV2}). The last term in (\ref{SC1-10}) which is
proportional to $\tilde G\eta_0$ does not give rise to a mixing of $\tilde G$ and $\eta_0$ because
it vanishes after radial integration. However, in the presence of metric fluctuations it no
longer vanishes and gives rise to the interaction terms in 
$\mathcal L_{\tilde{G}\eta_0 G}$ listed in Eq.~(\ref{LGTeG}). 

In order to determine the values of interaction vertices, we need to normalize the pseudoscalar
glueball fields.\0{}
Demanding that the pseudoscalar glueball fields $\tilde G^{(n)}(x)$ appearing in the mode expansion of
$C_\tau'$ have canonical kinetic terms
fixes the normalization of the radial mode functions through
\bea
&&\frac{\Vv \delta\tau}{2\pi(2\pi l_s)^6}\RSS^3\int_{\uKK}^\infty du\,u f^{-1}(u)[V^{(n)}(u)]^2\nonumber\\
&&=\frac{\lambda^3}{4\cdot 3^5 \pi^4}\int_1^\infty d\bar u\,\bar u f^{-1}(\bar u) [V^{(n)}(\bar u)]^2 = 1
\eea
with $\Vv=8\pi^2/3$ being the volume of the unit $S^4$.
For the lightest and the first excited pseudoscalar glueball mode this implies
\bea
&& [\bar V^{(1)}(\bar u=1)]^{-1}=0.002046\ldots \lambda^{3/2},\\ 
&& [\bar V^{(2)}(\bar u=1)]^{-1}=0.001157\ldots \lambda^{3/2}.
\eea

Using the mode expansions of the metric fields given in Ref.~\cite{Brunner:2015oqa},
the effective Lagrangian for pseudoscalar glueball interactions can be obtained
by numerical integrations over products of the relevant radial mode functions.

\bibliographystyle{JHEP}
\bibliography{glueballdecay}

\providecommand{\href}[2]{#2}\begingroup\raggedright\begin{thebibliography}{10}

\bibitem{Fritzsch:1972jv}
H.~Fritzsch and M.~Gell-Mann, {\it {Current algebra: Quarks and what else?}},
  {\em eConf} {\bf C720906V2} (1972) 135--165,
  [\href{http://arxiv.org/abs/hep-ph/0208010}{{\tt hep-ph/0208010}}].

\bibitem{Fritzsch:1975tx}
H.~Fritzsch and P.~Minkowski, {\it {$\Psi$ Resonances, Gluons and the Zweig
  Rule}},  {\em Nuovo Cim.} {\bf A30} (1975) 393.

\bibitem{Jaffe:1975fd}
R.~Jaffe and K.~Johnson, {\it {Unconventional States of Confined Quarks and
  Gluons}},  {\em Phys.Lett.} {\bf B60} (1976) 201.

\bibitem{Bugg:2004xu}
D.~Bugg, {\it {Four sorts of meson}},  {\em Phys.Rept.} {\bf 397} (2004)
  257--358, [\href{http://arxiv.org/abs/hep-ex/0412045}{{\tt hep-ex/0412045}}].

\bibitem{Klempt:2007cp}
E.~Klempt and A.~Zaitsev, {\it {Glueballs, Hybrids, Multiquarks. Experimental
  facts versus QCD inspired concepts}},  {\em Phys.Rept.} {\bf 454} (2007)
  1--202, [\href{http://arxiv.org/abs/0708.4016}{{\tt arXiv:0708.4016}}].

\bibitem{Crede:2008vw}
V.~Crede and C.~Meyer, {\it {The Experimental Status of Glueballs}},  {\em
  Prog.Part.Nucl.Phys.} {\bf 63} (2009) 74--116,
  [\href{http://arxiv.org/abs/0812.0600}{{\tt arXiv:0812.0600}}].

\bibitem{Ochs:2013gi}
W.~Ochs, {\it {The Status of Glueballs}},  {\em J.Phys.} {\bf G40} (2013)
  043001, [\href{http://arxiv.org/abs/1301.5183}{{\tt arXiv:1301.5183}}].

\bibitem{Donoghue:1980hw}
J.~F. Donoghue, K.~Johnson, and B.~A. Li, {\it {Low Mass Glueballs in the Meson
  Spectrum}},  {\em Phys. Lett.} {\bf B99} (1981) 416--420.

\bibitem{Ishikawa:1980xv}
K.~Ishikawa, {\it {Is the E(1420) in $J/\psi$ Decay a Gluonic Bound State?}},
  {\em Phys. Rev. Lett.} {\bf 46} (1981) 978.

\bibitem{Chanowitz:1980gu}
M.~S. Chanowitz, {\it {Have We Seen Our First Glueball?}},  {\em Phys. Rev.
  Lett.} {\bf 46} (1981) 981.

\bibitem{Edwards:1982nc}
C.~Edwards et~al., {\it {Identification of a Pseudoscalar State at 1440 MeV in
  $J/\psi$ Radiative Decays}},  {\em Phys. Rev. Lett.} {\bf 49} (1982) 259.
  [Erratum: Phys. Rev. Lett.50,219(1983)].

\bibitem{PDG16}
C.~Patrignani et~al., {\it {Review of Particle Physics}},  {\em Chin. Phys.}
  {\bf C40} (2016) 100001.

\bibitem{Masoni:2006rz}
A.~Masoni, C.~Cicalo, and G.~L. Usai, {\it {The case of the pseudoscalar
  glueball}},  {\em J. Phys.} {\bf G32} (2006) R293--R335.

\bibitem{Amsler:1995td}
C.~Amsler and F.~E. Close, {\it {Is $f_0(1500)$ a scalar glueball?}},  {\em
  Phys.Rev.} {\bf D53} (1996) 295--311,
  [\href{http://arxiv.org/abs/hep-ph/9507326}{{\tt hep-ph/9507326}}].

\bibitem{Lee:1999kv}
W.-J. Lee and D.~Weingarten, {\it {Scalar quarkonium masses and mixing with the
  lightest scalar glueball}},  {\em Phys.Rev.} {\bf D61} (1999) 014015,
  [\href{http://arxiv.org/abs/hep-lat/9910008}{{\tt hep-lat/9910008}}].

\bibitem{Close:2001ga}
F.~E. Close and A.~Kirk, {\it {Scalar glueball $q \bar q$ mixing above 1 GeV
  and implications for lattice QCD}},  {\em Eur.Phys.J.} {\bf C21} (2001)
  531--543, [\href{http://arxiv.org/abs/hep-ph/0103173}{{\tt hep-ph/0103173}}].

\bibitem{Amsler:2004ps}
C.~Amsler and N.~T{\"o}rnqvist, {\it {Mesons beyond the naive quark model}},
  {\em Phys.Rept.} {\bf 389} (2004) 61--117.

\bibitem{Close:2005vf}
F.~E. Close and Q.~Zhao, {\it {Production of $f_0(1710)$, $f_0(1500)$, and
  $f_0(1370)$ in $J/\psi$ hadronic decays}},  {\em Phys.Rev.} {\bf D71} (2005)
  094022, [\href{http://arxiv.org/abs/hep-ph/0504043}{{\tt hep-ph/0504043}}].

\bibitem{Giacosa:2005zt}
F.~Giacosa, T.~Gutsche, V.~Lyubovitskij, and A.~Faessler, {\it {Scalar nonet
  quarkonia and the scalar glueball: Mixing and decays in an effective chiral
  approach}},  {\em Phys.Rev.} {\bf D72} (2005) 094006,
  [\href{http://arxiv.org/abs/hep-ph/0509247}{{\tt hep-ph/0509247}}].

\bibitem{Albaladejo:2008qa}
M.~Albaladejo and J.~Oller, {\it {Identification of a Scalar Glueball}},  {\em
  Phys.Rev.Lett.} {\bf 101} (2008) 252002,
  [\href{http://arxiv.org/abs/0801.4929}{{\tt arXiv:0801.4929}}].

\bibitem{Mathieu:2008me}
V.~Mathieu, N.~Kochelev, and V.~Vento, {\it {The Physics of Glueballs}},  {\em
  Int.J.Mod.Phys.} {\bf E18} (2009) 1--49,
  [\href{http://arxiv.org/abs/0810.4453}{{\tt arXiv:0810.4453}}].

\bibitem{Janowski:2011gt}
S.~Janowski, D.~Parganlija, F.~Giacosa, and D.~H. Rischke, {\it {The Glueball
  in a Chiral Linear Sigma Model with Vector Mesons}},  {\em Phys.Rev.} {\bf
  D84} (2011) 054007, [\href{http://arxiv.org/abs/1103.3238}{{\tt
  arXiv:1103.3238}}].

\bibitem{Janowski:2014ppa}
S.~Janowski, F.~Giacosa, and D.~H. Rischke, {\it {Is $f_0(1710)$ a glueball?}},
   {\em Phys.Rev.} {\bf D90} (2014) 114005,
  [\href{http://arxiv.org/abs/1408.4921}{{\tt arXiv:1408.4921}}].

\bibitem{Cheng:2015iaa}
H.-Y. Cheng, C.-K. Chua, and K.-F. Liu, {\it {Revisiting Scalar Glueballs}},
  {\em Phys. Rev.} {\bf D92} (2015) 094006,
  [\href{http://arxiv.org/abs/1503.06827}{{\tt arXiv:1503.06827}}].

\bibitem{Close:2015rza}
F.~E. Close and A.~Kirk, {\it {Interpretation of scalar and axial mesons in
  LHCb from a historical perspective}},  {\em Phys. Rev.} {\bf D91} (2015)
  114015, [\href{http://arxiv.org/abs/1503.06942}{{\tt arXiv:1503.06942}}].

\bibitem{Frere:2015xxa}
J.-M. Fr{\`e}re and J.~Heeck, {\it {Scalar glueballs: Constraints from the
  decays into $\eta$ or $\eta'$}},  {\em Phys. Rev.} {\bf D92} (2015), no.~11
  114035, [\href{http://arxiv.org/abs/1506.04766}{{\tt arXiv:1506.04766}}].

\bibitem{Morningstar:1999rf}
C.~J. Morningstar and M.~J. Peardon, {\it {The Glueball spectrum from an
  anisotropic lattice study}},  {\em Phys.Rev.} {\bf D60} (1999) 034509,
  [\href{http://arxiv.org/abs/hep-lat/9901004}{{\tt hep-lat/9901004}}].

\bibitem{Chen:2005mg}
Y.~Chen, A.~Alexandru, S.~Dong, T.~Draper, I.~Horvath, et~al., {\it {Glueball
  spectrum and matrix elements on anisotropic lattices}},  {\em Phys.Rev.} {\bf
  D73} (2006) 014516, [\href{http://arxiv.org/abs/hep-lat/0510074}{{\tt
  hep-lat/0510074}}].

\bibitem{Gabadadze:1997zc}
G.~Gabadadze, {\it {Pseudoscalar glueball mass: QCD versus lattice gauge theory
  prediction}},  {\em Phys. Rev.} {\bf D58} (1998) 055003,
  [\href{http://arxiv.org/abs/hep-ph/9711380}{{\tt hep-ph/9711380}}].

\bibitem{Richards:2010ck}
{\bf UKQCD} Collaboration, C.~M. Richards, A.~C. Irving, E.~B. Gregory, and
  C.~McNeile, {\it {Glueball mass measurements from improved staggered fermion
  simulations}},  {\em Phys. Rev.} {\bf D82} (2010) 034501,
  [\href{http://arxiv.org/abs/1005.2473}{{\tt arXiv:1005.2473}}].

\bibitem{Gregory:2012hu}
E.~Gregory, A.~Irving, B.~Lucini, C.~McNeile, A.~Rago, et~al., {\it {Towards
  the glueball spectrum from unquenched lattice QCD}},  {\em JHEP} {\bf 1210}
  (2012) 170, [\href{http://arxiv.org/abs/1208.1858}{{\tt arXiv:1208.1858}}].

\bibitem{Sun:2017ipk}
W.~Sun, L.-C. Gui, Y.~Chen, M.~Gong, C.~Liu, Y.-B. Liu, Z.~Liu, J.-P. Ma, and
  J.-B. Zhang, {\it {Glueball spectrum from $N_f=2$ lattice QCD study on
  anisotropic lattices}},  \href{http://arxiv.org/abs/1702.08174}{{\tt
  arXiv:1702.08174}}.

\bibitem{Bugg:2009cf}
D.~V. Bugg, {\it {Data on $J/\Psi \to \gamma(K^{\pm} K^0_S \pi^{\mp})$ and
  $\gamma(\eta \pi^+ \pi^-)$}},  \href{http://arxiv.org/abs/0907.3015}{{\tt
  arXiv:0907.3015}}.

\bibitem{Cheng:2008ss}
H.-Y. Cheng, H.-n. Li, and K.-F. Liu, {\it {Pseudoscalar glueball mass from
  $\eta$-$\eta'$-$G$ mixing}},  {\em Phys. Rev.} {\bf D79} (2009) 014024,
  [\href{http://arxiv.org/abs/0811.2577}{{\tt arXiv:0811.2577}}].

\bibitem{Ambrosino:2009sc}
{\bf KLOE} Collaboration, F.~Ambrosino et~al., {\it {A Global fit to determine
  the pseudoscalar mixing angle and the gluonium content of the $\eta'$
  meson}},  {\em JHEP} {\bf 0907} (2009) 105,
  [\href{http://arxiv.org/abs/0906.3819}{{\tt arXiv:0906.3819}}].

\bibitem{Gutsche:2009jh}
T.~Gutsche, V.~E. Lyubovitskij, and M.~C. Tichy, {\it {$\eta(1405)$ in a chiral
  approach based on mixing of the pseudoscalar glueball with the first radial
  excitations of $\eta$ and $\eta'$}},  {\em Phys. Rev.} {\bf D80} (2009)
  014014, [\href{http://arxiv.org/abs/0904.3414}{{\tt arXiv:0904.3414}}].

\bibitem{Eshraim:2012jv}
W.~I. Eshraim, S.~Janowski, F.~Giacosa, and D.~H. Rischke, {\it {Decay of the
  pseudoscalar glueball into scalar and pseudoscalar mesons}},  {\em Phys.
  Rev.} {\bf D87} (2013) 054036, [\href{http://arxiv.org/abs/1208.6474}{{\tt
  arXiv:1208.6474}}].

\bibitem{Eshraim:2016mds}
W.~I. Eshraim and S.~Schramm, {\it {Decay modes of the excited pseudoscalar
  glueball}},  {\em Phys. Rev.} {\bf D95} (2017) 014028,
  [\href{http://arxiv.org/abs/1606.02207}{{\tt arXiv:1606.02207}}].

\bibitem{Rebhan:2014rxa}
A.~Rebhan, {\it {The Witten-Sakai-Sugimoto model: A brief review and some
  recent results}},  {\em EPJ Web Conf.} {\bf 95} (2015) 02005,
  [\href{http://arxiv.org/abs/1410.8858}{{\tt arXiv:1410.8858}}].

\bibitem{Brunner:2015oqa}
F.~Br{\"u}nner, D.~Parganlija, and A.~Rebhan, {\it {Glueball Decay Rates in the
  Witten-Sakai-Sugimoto Model}},  {\em Phys. Rev.} {\bf D91} (2015) 106002,
  [\href{http://arxiv.org/abs/1501.07906}{{\tt arXiv:1501.07906}}].

\bibitem{1504.05815}
F.~Br{\"u}nner and A.~Rebhan, {\it {Nonchiral enhancement of scalar glueball
  decay in the Witten-Sakai-Sugimoto model}},  {\em Phys. Rev. Lett.} {\bf 115}
  (2015) 131601, [\href{http://arxiv.org/abs/1504.05815}{{\tt
  arXiv:1504.05815}}].

\bibitem{1510.07605}
F.~Br{\"u}nner and A.~Rebhan, {\it {Constraints on the $\eta\eta'$ decay rate
  of a scalar glueball from gauge/gravity duality}},  {\em Phys. Rev.} {\bf
  D92} (2015) 121902, [\href{http://arxiv.org/abs/1510.07605}{{\tt
  arXiv:1510.07605}}].

\bibitem{Witten:1998zw}
E.~Witten, {\it {Anti-de Sitter space, thermal phase transition, and
  confinement in gauge theories}},  {\em Adv.Theor.Math.Phys.} {\bf 2} (1998)
  505--532, [\href{http://arxiv.org/abs/hep-th/9803131}{{\tt hep-th/9803131}}].

\bibitem{Sakai:2004cn}
T.~Sakai and S.~Sugimoto, {\it {Low energy hadron physics in holographic QCD}},
   {\em Prog.Theor.Phys.} {\bf 113} (2005) 843--882,
  [\href{http://arxiv.org/abs/hep-th/0412141}{{\tt hep-th/0412141}}].

\bibitem{Sakai:2005yt}
T.~Sakai and S.~Sugimoto, {\it {More on a holographic dual of QCD}},  {\em
  Prog.Theor.Phys.} {\bf 114} (2005) 1083--1118,
  [\href{http://arxiv.org/abs/hep-th/0507073}{{\tt hep-th/0507073}}].

\bibitem{Brower:2000rp}
R.~C. Brower, S.~D. Mathur, and C.-I. Tan, {\it {Glueball spectrum for QCD from
  AdS supergravity duality}},  {\em Nucl.Phys.} {\bf B587} (2000) 249--276,
  [\href{http://arxiv.org/abs/hep-th/0003115}{{\tt hep-th/0003115}}].

\bibitem{Bali:2013kia}
G.~S. Bali, F.~Bursa, L.~Castagnini, S.~Collins, L.~Del~Debbio, et~al., {\it
  {Mesons in large-N QCD}},  {\em JHEP} {\bf 1306} (2013) 071,
  [\href{http://arxiv.org/abs/1304.4437}{{\tt arXiv:1304.4437}}].

\bibitem{Hashimoto:2007ze}
K.~Hashimoto, C.-I. Tan, and S.~Terashima, {\it {Glueball decay in holographic
  QCD}},  {\em Phys.Rev.} {\bf D77} (2008) 086001,
  [\href{http://arxiv.org/abs/0709.2208}{{\tt arXiv:0709.2208}}].

\bibitem{Armoni:2004dc}
A.~Armoni, {\it {Witten-Veneziano from Green-Schwarz}},  {\em JHEP} {\bf 0406}
  (2004) 019, [\href{http://arxiv.org/abs/hep-th/0404248}{{\tt
  hep-th/0404248}}].

\bibitem{Barbon:2004dq}
J.~L. Barbon, C.~Hoyos-Badajoz, D.~Mateos, and R.~C. Myers, {\it {The
  Holographic life of the $\eta'$}},  {\em JHEP} {\bf 0410} (2004) 029,
  [\href{http://arxiv.org/abs/hep-th/0404260}{{\tt hep-th/0404260}}].

\bibitem{Arean:2016hcs}
D.~Are{\'a}n, I.~Iatrakis, M.~J{\"a}rvinen, and E.~Kiritsis, {\it {The CP-odd
  sector and $\theta$ dynamics in holographic QCD}},
  \href{http://arxiv.org/abs/1609.08922}{{\tt arXiv:1609.08922}}.

\bibitem{Anderson:2014jia}
N.~Anderson, S.~K. Domokos, J.~A. Harvey, and N.~Mann, {\it {Central production
  of $\eta$ and $\eta'$ via double Pomeron exchange in the Sakai-Sugimoto
  model}},  {\em Phys. Rev.} {\bf D90} (2014) 086010,
  [\href{http://arxiv.org/abs/1406.7010}{{\tt arXiv:1406.7010}}].

\bibitem{Rosenzweig:1981cu}
C.~Rosenzweig, A.~Salomone, and J.~Schechter, {\it {A Pseudoscalar Glueball,
  the Axial Anomaly and the Mixing Problem for Pseudoscalar Mesons}},  {\em
  Phys. Rev.} {\bf D24} (1981) 2545--2548.

\bibitem{Rosenzweig:1982cb}
C.~Rosenzweig, A.~Salomone, and J.~Schechter, {\it {How does a pseudoscalar
  glueball come unglued?}},  {\em Nucl.Phys.} {\bf B206} (1982) 12.

\bibitem{Rosenzweig:1979ay}
C.~Rosenzweig, J.~Schechter, and C.~G. Trahern, {\it {Is the Effective
  Lagrangian for QCD a Sigma Model?}},  {\em Phys. Rev.} {\bf D21} (1980) 3388.

\bibitem{Gounaris:1988rp}
G.~J. Gounaris and H.~Neufeld, {\it {Why $\iota(1460)$ decays mainly into $K
  \bar K \pi$?}},  {\em Phys. Lett.} {\bf B213} (1988) 541. [Erratum: Phys.
  Lett.B218,508(1989)].

\bibitem{Aharony:2008an}
O.~Aharony and D.~Kutasov, {\it {Holographic Duals of Long Open Strings}},
  {\em Phys.Rev.} {\bf D78} (2008) 026005,
  [\href{http://arxiv.org/abs/0803.3547}{{\tt arXiv:0803.3547}}].

\bibitem{Hashimoto:2008sr}
K.~Hashimoto, T.~Hirayama, F.-L. Lin, and H.-U. Yee, {\it {Quark Mass
  Deformation of Holographic Massless QCD}},  {\em JHEP} {\bf 0807} (2008) 089,
  [\href{http://arxiv.org/abs/0803.4192}{{\tt arXiv:0803.4192}}].

\bibitem{McNees:2008km}
R.~McNees, R.~C. Myers, and A.~Sinha, {\it {On quark masses in holographic
  QCD}},  {\em JHEP} {\bf 0811} (2008) 056,
  [\href{http://arxiv.org/abs/0807.5127}{{\tt arXiv:0807.5127}}].

\bibitem{0708.2839}
O.~Bergman, S.~Seki, and J.~Sonnenschein, {\it {Quark mass and condensate in
  HQCD}},  {\em JHEP} {\bf 0712} (2007) 037,
  [\href{http://arxiv.org/abs/0708.2839}{{\tt arXiv:0708.2839}}].

\bibitem{Dhar:2007bz}
A.~Dhar and P.~Nag, {\it {Sakai-Sugimoto model, Tachyon Condensation and Chiral
  symmetry Breaking}},  {\em JHEP} {\bf 0801} (2008) 055,
  [\href{http://arxiv.org/abs/0708.3233}{{\tt arXiv:0708.3233}}].

\bibitem{Dhar:2008um}
A.~Dhar and P.~Nag, {\it {Tachyon condensation and quark mass in modified
  Sakai-Sugimoto model}},  {\em Phys.Rev.} {\bf D78} (2008) 066021,
  [\href{http://arxiv.org/abs/0804.4807}{{\tt arXiv:0804.4807}}].

\bibitem{Niarchos:2010ki}
V.~Niarchos, {\it {Hairpin-Branes and Tachyon-Paperclips in Holographic
  Backgrounds}},  {\em Nucl. Phys.} {\bf B841} (2010) 268--302,
  [\href{http://arxiv.org/abs/1005.1650}{{\tt arXiv:1005.1650}}].

\bibitem{Lutz:2009ff}
M.~Lutz et~al., {\it {Physics Performance Report for PANDA: Strong Interaction
  Studies with Antiprotons}},  \href{http://arxiv.org/abs/0903.3905}{{\tt
  arXiv:0903.3905}}.

\bibitem{Eshraim:2012rb}
W.~Eshraim, S.~Janowski, A.~Peters, K.~Neuschwander, and F.~Giacosa, {\it
  {Interaction of the pseudoscalar glueball with (pseudo)scalar mesons and
  nucleons}},  {\em Acta Phys. Polon. Supp.} {\bf 5} (2012) 1101--1108,
  [\href{http://arxiv.org/abs/1209.3976}{{\tt arXiv:1209.3976}}].

\bibitem{Herzog:2008mu}
C.~P. Herzog, S.~Paik, M.~J. Strassler, and E.~G. Thompson, {\it {Holographic
  Double Diffractive Scattering}},  {\em JHEP} {\bf 08} (2008) 010,
  [\href{http://arxiv.org/abs/0806.0181}{{\tt arXiv:0806.0181}}].

\bibitem{Brower:2012mk}
R.~C. Brower, M.~Djuri{\'c}, and C.-I. Tan, {\it {Diffractive Higgs Production
  by AdS Pomeron Fusion}},  {\em JHEP} {\bf 09} (2012) 097,
  [\href{http://arxiv.org/abs/1202.4953}{{\tt arXiv:1202.4953}}].

\bibitem{Harland-Lang:2013ncy}
L.~A. Harland-Lang, V.~A. Khoze, M.~G. Ryskin, and W.~J. Stirling, {\it
  {Central exclusive production as a probe of the gluonic component of the
  $\eta'$ and $\eta$ mesons}},  {\em Eur. Phys. J.} {\bf C73} (2013) 2429,
  [\href{http://arxiv.org/abs/1302.2004}{{\tt arXiv:1302.2004}}].

\bibitem{Close:1997pj}
F.~E. Close and A.~Kirk, {\it {A glueball - $q\bar q$ filter in central hadron
  production}},  {\em Phys. Lett.} {\bf B397} (1997) 333--338,
  [\href{http://arxiv.org/abs/hep-ph/9701222}{{\tt hep-ph/9701222}}].

\bibitem{Kirk:2014nwa}
A.~Kirk, {\it {A review of central production experiments at the CERN Omega
  spectrometer}},  {\em Int. J. Mod. Phys.} {\bf A29} (2014) 1446001,
  [\href{http://arxiv.org/abs/1408.1196}{{\tt arXiv:1408.1196}}].

\bibitem{Kruczenski:2003uq}
M.~Kruczenski, D.~Mateos, R.~C. Myers, and D.~J. Winters, {\it {Towards a
  holographic dual of large $N_c$ QCD}},  {\em JHEP} {\bf 0405} (2004) 041,
  [\href{http://arxiv.org/abs/hep-th/0311270}{{\tt hep-th/0311270}}].

\bibitem{Kanitscheider:2008kd}
I.~Kanitscheider, K.~Skenderis, and M.~Taylor, {\it {Precision holography for
  non-conformal branes}},  {\em JHEP} {\bf 0809} (2008) 094,
  [\href{http://arxiv.org/abs/0807.3324}{{\tt arXiv:0807.3324}}].

\bibitem{Bigazzi:2015bna}
F.~Bigazzi, A.~L. Cotrone, and R.~Sisca, {\it {Notes on Theta Dependence in
  Holographic Yang-Mills}},  {\em JHEP} {\bf 08} (2015) 090,
  [\href{http://arxiv.org/abs/1506.03826}{{\tt arXiv:1506.03826}}].

\end{thebibliography}\endgroup

\end{document}